\documentstyle[aps,prb,epsf,multicol]{revtex}

\newcommand{\ffr}{f_{\!f\!r}} \newcommand{\Ffr}{F_{\!f\!r}}
\newcommand{\hb}{\bar{h}} \newcommand{\lb}{\bar{l}}
\newcommand{\bx}{{\bf x}} 
\newcommand{\bk}{{\bf k}}
\newcommand{\bq}{{\bf q}}
\newcommand{\bp}{{\bf p}}
\newcommand{\fq}{{\vec\xi}}
\newcommand{\fp}{{\vec\chi}}
\newcommand{\br}{{\bf r}}
\newcommand{\E}{{\rm e}}
\newcommand{\I}{{\rm i}}
\newcommand{\Dx}{d^D\!\!x}
 
\newcommand{\bea}{\begin{eqnarray}}
\newcommand{\eea}{\end{eqnarray}}
\newcommand{\be}{\begin{equation}}
\newcommand{\ee}{\end{equation}}
\newcommand{\lessim}{\raisebox{-0.5ex}{$\;\stackrel{<}{\scriptstyle\sim}\;$}}
\newcommand{\morsim}{\raisebox{-0.5ex}{$\;\stackrel{>}{\scriptstyle\sim}\;$}}
\newcommand{\DS}{\displaystyle}

\begin{document}
\draft
\title{Towards a statistical theory of solid dry friction}
\author{\rm Andreas Volmer\thanks{e-mail {\tt av@thp.uni-koeln.de}}
 and Thomas Nattermann}
\address{Institut f\"ur
  theoretische Physik, Universit\"at K\"oln, Z\"ulpicher Str. 77,
  D-50937 K\"oln}
\maketitle

\begin{abstract}
  Wearless dry friction of an elastic block of weight $N$, driven by
  an external force $F$ over a rigid substrate, is investigated. The
  slider and substrate surfaces are both microscopically rough,
  interacting via a repulsive potential that depends on the local
  overlap. The model reproduces Amontons's laws which state that the
  friction force is proportional to the normal loading force $N$ and
  independent of the nominal surface area. In this model, the dynamic
  friction force decays for large velocities and approaches a finite
  static friction for small velocities if the surface profiles are
  self-affine on small length scales.
\end{abstract}

\begin{multicols}{2}

\section{Introduction}

The physics of solid dry friction is an old and fascinating field.
Yet, many quite fundamental problems are still subject of debate. The
basic phenomenological facts, though, have since long been known as
the Coulomb-Amontons's laws of friction: (i) The frictional force is
independent of the size of the surfaces in contact, (ii) friction is
proportional to the normal load, and (iii) kinetic friction is not (or
not much) dependent on the velocity and typically lower than the
static friction force \cite{Bowden}.

A simple explanation for these laws arises from Bowden and Tabor's
adhesion theory of friction, in which plastic deformation of the
surfaces accounts for the load dependence of real contact area and
friction force \cite{Bowden,Scholz}. Furthermore, plastic deformation
leads to a logarithmic time dependence of the static friction and
logarithmic velocity dependence of the kinetic friction
\cite{Dieterich,HeslotEtAl}. Although plastic flow is assumed to yield
the main contribution to solid friction, other mechanisms may play a
role as well. In particular, it was noticed long ago that elastical
multistability and hysteresis also gives rise to friction
\cite{Tomlinson}.

Recently, steps have been taken towards the understanding of wearless
friction as a collective phenomenon, dominated by the competition of
pinning forces emerging from rough surfaces and bulk elasticity,
neglecting plastic deformations \cite{CaroliNozieres,PerssonTossati}.
However, a quantitative understanding is still lacking. Attempts in
this direction are inspired by studies of the depinning transition of
driven charge density waves \cite{EfetovLarkin77}, interfaces in
random media \cite{KopLev,NatterLeschhorn} and of vortex lines in
type-II superconductors \cite{Blatter}, where the behaviour near the
depinning threshold force turned out to be a non-equilibrium critical
phenomenon described by new universal critical exponents and scaling
laws. It is tempting to assume that friction is a related phenomenon.
Indeed, in a recent investigation by Cule and Hwa \cite{CuleHwa}, a
bead-spring model for friction has been considered which exhibits a
depinning transition of the universality class of interface depinning.
Bead or block chain models do however not account for Amontons's laws.
The aim of the present paper is to study a simple statistical model
where friction solely arises from hysteretic elastic response, and to
find whether it is nevertheless capable of reproducing these
fundamental laws.  The situation we consider is the weak pinning
limit, where elastic multistability arises as a collective effect. The
opposite limit of strong pinning, where multistability already emerges
on the local scale of single traps, has been considered recently by
Caroli and Nozi\`eres \cite{CaroliNozieres}.

\section{The Model}

\begin{figure}[b]
  \begin{center}
    \leavevmode \epsfxsize=0.4\textwidth \epsfbox{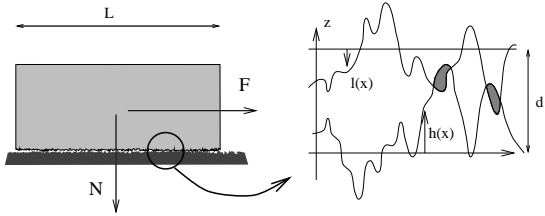}
  \end{center}
\narrowtext
\caption{\label{sketch} Cartoon of the model. A block of weight $N$ is
pulled over a rigid substrate. The two adjacent surface profiles $h$
and $l$, separated by a mean distance $d$, are enlarged. Typical
overlap areas are shown.} 
\end{figure}

To be specific, we consider an elastic body of weight $N$ and linear
size $L$, which is pulled over a rigid substrate, cf.\ 
Fig.~\ref{sketch}. The two surface profiles, separated by a mean
distance $d$, are parameterized by scalar height functions $l(\bx)$
and $h(\bf x)$, respectively. $\bx$ denotes the 
2-dimensional position vector in the reference plane parallel to the
surface and the substrate. For simplicity, both surfaces are assumed
to have the same statistical properties: They independently obey
Gaussian distributions with mean zero, characterized by a short-range
pair correlation function 
\begin{equation}
\label{def_kx}
  \langle h(\bx)h(\bx') \rangle
  \equiv \hb^2 \,k((\bx-\bx')/\sigma),
\end{equation}
where $\hb$ defines the width of the substrate surface, $\sigma$ is
the typical lateral corrugation length, $k({\bf 0})=1$ and
$k(\bx)\approx 0$ for $|\bx|\gg 1$.  
Correspondingly, $\lb^2\equiv\langle l^2(0)\rangle$ describes the 
width of the slider surface. Unless otherwise stated,
we will however assume throughout the paper 
that both surface profiles obey the same distribution, hence
$\lb=\hb$. 

Short range correlations characterize a macroscopically flat surface.
Self-affine surfaces, on the other hand, are characterized, in Fourier
space, by a height-height correlator $\langle \tilde h_k \tilde h_{k'}
\rangle \;\sim\; \delta({\bf k+k'}) |\bk|^{-2\zeta-2}$, where
$\zeta$ is the roughness exponent. Fracture surfaces, for instance,
typically have $\zeta=0.6\ldots 0.9$ \cite{Zhang}. The power law
behaviour is usually cut off below some wave number $1/\sigma$
\cite{Scholz}, so it is reasonable to restrict the study first to
short-range correlations.

The lateral elastic properties of the bottom slider surface depend on
the shape of the body and are usually quite complicated.  In general,
the elastic energy can be written as \cite{Landau7}
\begin{equation}
  \label{GenElast}
  \int\!\!d^2x\!\int\!\!d^2x' \gamma_{\alpha\beta\gamma\delta}({\bf
    x-x'})\partial_{\alpha} \br_{\beta}(\bx)
  \partial'_{\gamma} \br_{\delta}(\bx'),
\end{equation}
where $\br(\bx)$ denotes the local lateral displacement from
the equilibrium position.  For simplicity, we restrict ourselves to
$\gamma_{\alpha\beta\gamma\delta}(\bx) = \gamma(\bx)\delta_{\alpha\gamma}
\delta_{\beta\delta}$ and consider only two limiting cases: If the
slider is a 2-dimensional object, like a latex membrane pulled over a
rod \cite{ValletteGollub}, $\gamma(\bx)=\gamma \delta(\bx)$,
i.e.\ the elastic interaction is local.
If, on the other hand, the slider is a semi-infinite 3-dimensional
object, elastic response is mediated by bulk elasticity and nonlocal
in space with $\gamma(\bx)\approx\gamma/|\bx|$.  Both cases
can be treated simultaneously by introducing an exponent $\alpha$,
so that the elastic kernel in Fourier space scales like
\begin{equation}
  \tilde\gamma(\bk)\equiv\int d^2 x \,\E^{\I\bk\bx} \gamma(\bx) \sim
  k^{\alpha-2}, 
\end{equation}
with $\alpha=1$ and 2 for bulk and surface elasticity,
respectively.  Consequently, the dispersion of the elastic energy will
behave as $k^\alpha$. Generalizing Eq.~(\ref{GenElast}) to a
$D$-dimensional surface, the real space elastic kernel scales like
$\gamma(\bx)\sim |\bx|^{2-D-\alpha}$. 

In the direction perpendicular to the reference plane, the slider
surface would strictly also have to be treated as elastic, interacting
with the substrate via a hard wall potential (in the absence of
adhesion forces). To make the model analytically amenable, however, we
allow the surfaces to overlap and introduce a repulsive potential
$V(z)$ that depends on the local overlap 
\begin{equation}
  z(\bx) = h(\bx+{\bf r}(\bx,t))+l(\bx)-d.
\end{equation}
This potential is used to mimic vertical elasticity. 
We choose 
\begin{equation}
  \label{Vdef}
  V(z) = V_0\,z^n\,\Theta(z),
\end{equation}
where
$\Theta(z)$ is the Heaviside step function, and $n>1$. With $n=3/2$,
the Hertzian result on the distance dependence of the repulsive force
between two elastic spheres \cite{Landau7} is reproduced (this
nontrivial dependence being the result of the interplay between
Hooke's law and the spherical geometry). As it turns out, the
results do not depend very sensitively on the chosen value of
$n$. Note that setting $V(z)\equiv 0$ for $z<0$ via the Heaviside
function is natural and justified, in the absence of adhesion. 

Finally, the total driving force ${\bf F}$ is applied
homogeneously as a force density ${\bf f = F}/L^2$. This choice
appears to be natural in the case of bulk elasticity, whilst for a
membrane-like slider the external force should rather be exerted at
one border.

Assume that, with the slider lying at rest on the substrate, we turn
on the external driving force ${\bf F}$. If $F$ is large enough, the
slider will start to move, and the interaction between the slider and
the substrate will generate a dynamic friction force $\Ffr\equiv -F$
(as we will show below) which leads, after a transient time of
acceleration, to a constant average velocity ${\bf v}$ of the center
of mass of the slider. In the steady state, the equation of motion of
a point $\br(\bx,t)$ of the slider surface can be written as
\begin{eqnarray}
  \label{u_eom}
  \eta \,(\dot\br(\bx,t) -{\bf v}) &=& \int d^2x'\gamma({\bf
    x-x'}) \nabla'^2 \br({\bf x'},t) + {\bf f}  \\
  &&  -\frac{\partial}{\partial \br(\bx,t)} V[h(\bx+{\bf
    r}(\bx,t))+l(\bx)-d]. \nonumber
\end{eqnarray}
The term on the left hand side accounts for surface phononic damping
{\em within} the slider, with a damping coefficient $\eta$. Since
Eq.~(\ref{u_eom}) is written in the laboratory frame, the center of
mass velocity ${\bf v}$ has to be subtracted from the local velocity
$\dot\br$. In this way we make sure that rigid sliding, i.e.\
$\br(\bx)\equiv {\bf v}t$ with a constant velocity ${\bf v}$, is
frictionless. 
The neglect of an inertial term is justified by our
primary interest in the depinning region, where kinetic energy is
relatively small.  If we started, on the other hand, with a theory
including an inertial term only, viscous friction would nevertheless
be generated by the nonlinear random force after eliminating short
wavelength displacement modes. 

The mean separation $d$ between slider and substrate obeys a similar
equation of motion
\begin{equation}
  \label{d_eom}
  \Lambda \frac{\partial}{\partial t}d(t) = -N - \int
  d^2x\;\frac{\partial}{\partial d} V[h(\bx+\br({\bf
    x},t))+l(\bx)-d],
\end{equation}
which has to be solved simultaneously with Eq.~(\ref{u_eom}).  Again,
overdamped motion is assumed with another friction constant. In the
analysis we will take $\Lambda\rightarrow\infty$, because in the
thermodynamic limit of infinitely large system size $L$, fluctuations
of the mean distance $d$ will vanish, so that it can be treated as a
constant parameter that has to be determined self-consistently.

Simple scaling analysis shows that the model is dominated by two
dimensionless quantities:
\begin{equation}
\label{EN_def}
{\cal E}\equiv V_0\hb^n/\gamma\sigma^{2-\alpha} \qquad\mbox{and}\qquad
{\cal N}\equiv N/V_0\,\hb^{n-1}\sigma^2,
\end{equation}
where ${\cal E}$ is the ratio between typical overlap and elastic
forces on short scales. In this paper, we will restrict ourselves to
${\cal E}\ll 1$, corresponding to weak disorder or weak pinning, and
$1\ll {\cal N}\ll (L/\sigma)^2$. Under these conditions, ${\cal N}$ will
turn out to be of the order of the number of contact points between
the two surfaces ($V_0\,\hb^{n-1}\sigma^2$ is the typical overlap
force at a single contact), hence the latter condition ensures that
the real contact area is smaller than the nominal area $L^2$ and
bigger than the typical size $\sigma^2$ of one point of contact. Note
that ${\cal N}\ll (L/\sigma)^2$ can be rewritten as $p\ll p_0$, where
$p=N/L^2$ is the nominal pressure and $p_0=V_0\hb^{n-1}$ a typical
pressure at contact points.

\section{Static properties}

Let us first consider the limit of vanishing surface elasticity ${\cal
  E}\rightarrow 0$ (which implies that there is no friction force and
hence no pinning at all) and zero external force ${\bf F}$. In this
limit $\br (\bx,t)\equiv 0$, and in Eq.~(\ref{d_eom}) the
height profiles $h$ and $l$ can be averaged over. We find
\begin{eqnarray}
\label{NormalLoad}
  {\cal N} &=& \frac{1}{V_0\hb^{n-1}\sigma^2}
  \int d^2x \;\langle V'[h(\bx)+l(\bx)-d] \rangle \\
&=& c_n \, \left(\frac{L}{\sigma}\right)^2 \, (2\hb/d)^n
\E^{-d^2/4\hb^2} \;\left(1+{\cal O}\left((\hb/d)^2\right)\right),
\end{eqnarray}
where $c_n\equiv \Gamma(n+1)/\sqrt{4\pi}$. This implicitly determines
$d$ as a function of ${\cal N}$ and the surface roughness: $d\approx
2\hb \ln^{1/2} (c_n L^2/{\cal N}\sigma^2)$.  The real contact area
\begin{equation}
  \label{def_Ar}
  A_r\equiv L^2\langle \Theta(h(\bx)+l(\bx)-d)\rangle
\end{equation}
can be calculated in the same limit. Note that, while lateral
elasticity is suppressed in the limit ${\cal E}\rightarrow 0$, 
the "soft" interaction potential $V(z)$ that mimics vertical
elasticity still 
allows for a finite contact area. Making use of
Eq.~(\ref{NormalLoad}), it is given by 
\begin{equation}
\label{Coverage}
A_r \approx \frac{{\cal N}\sigma^2}{\sqrt{4\pi}\,c_n}
\ln^{\frac{4-n}{2n}}\left[c_nL^2/{\cal N}\sigma^2 \right] .
\end{equation}
Apart from logarithmic corrections, $A_r$ is thus proportional to the
normal load, independent of the total surface area\footnote{In
  \cite{Greenwood}, the proportionality between load and contact area
  was shown for a purely elastic surface with Gaussian distributed
  profile. In the model considered there, surface asperities were
  assumed to be spherically shaped and to obey the Hertzian
  distance-force relation.}. The dependence on the potential
parameter $n$ is weak, influencing only the strength of the
logarithmic correction.

The proportionality between load and contact area is a generic feature
in our model, as long as the height probability distribution $P(h)$
decreases at least exponentially for heights $|h|\gg\hb$, where $\hb$
is the surface profile width. The only restriction imposed on 
the choice of the potential $V(z)$ is that it is cut off for $z<0$,
which should be true for any effective, adhesionless interaction, and
that it increases slower than exponentially, or, in the case of a
Gaussian height probability distribution, at most exponentially; the
exact form for $z>0$ does not matter. Under these assumptions, the
expectation value (considered to be a function of $d$) satisfies
\begin{equation}
  \langle V(h-d) \rangle \equiv 
  \int_d^\infty V(h-d)P(h)dh
  \;\sim\; P(d)
\end{equation}
to leading order. With the definition in Eq.~(\ref{def_Ar}) and
$N/L^2=\langle V'(h+l-d)\rangle$, the proportionality between $A_r$ and
$N$ is a consequence of this result.

We next assume that the elasticity constant ${\cal E}$ is finite and
set up a perturbation expansion in the local displacement field ${\bf
  r}(\bx,t)$ about an undistorted configuration, first with $F=0$.
For our further calculation it is useful to generalize the model to a
$D$-dimensional slider surface, redefining ${\cal N}$ correspondingly
to ${\cal N}=N/V_0\hb^{n-1}\sigma^D$. The definition of ${\cal E}$ is
unaffected. The set of equations of motion (\ref{u_eom}) and
(\ref{d_eom}) can 
then be solved by iteration, order by order in the strength of the
interaction potential $V_0$, using a diagrammatic technique introduced
in \cite{EfetovLarkin74}. At each step, the random profiles $h$ and
$l$ are averaged over. In the following analysis, the mean distance
is assumed to be determined self-consistently by Eq.~(\ref{d_eom}) in the limit
$\Lambda\rightarrow\infty$, i.e. it is treated as a constant.

In order to set up the perturbation theory, we transform
the equation of motion (\ref{u_eom}) to Fourier space:
\begin{eqnarray}
  \label{r_eom_fourier}
  &&G_0^{-1}(\bk,\omega)\; \tilde \br_{\bk,\omega} =
  (2\pi)^{D+1}\delta^{D}(\bk)\delta(\omega) {\bf f} \\ 
  &&-\int\!\!
  \Dx\,dt\,\E^{-\I\bk\bx -\I\omega t} \frac{\partial}{\partial \br(\bx,t)}
  V[h(\bx+\br(\bx,t))+l(\bx)-d], \nonumber
\end{eqnarray}
where the bare propagator is given by
\begin{equation}
  G_0(\bk,\omega) = (\gamma|\bk|^\alpha +\I\eta \omega)^{-1}.
\end{equation}

Expanding the last term in Eq.~(\ref{r_eom_fourier}) in powers of
$\br(\bx,t)$, one gets for the displacement field to first order in
$V_0$
\begin{eqnarray}
 \label{r_firstorder}
 \tilde \br_{\bk,\omega} &=& -G_0(\bk,\omega) \int\Dx\,dt\,\E^{-\I\bk\bx
   -\I\omega t} \times \\
  && \qquad\qquad\nabla h(\bx)\,V'[h(\bx)+l(\bx)-d] + {\cal
   O}(V_0^2). \nonumber 
\end{eqnarray}
For the displacement correlation function in lowest non-vanishing
order follows
\begin{eqnarray}
  \label{paircorr}
  \langle \tilde \br_{\bk,\omega}\tilde \br_{\bk',\omega'}\rangle &=&
  -(2\pi)^{D+2}\delta(\omega)\,\delta(\omega')\delta^D(\bk+\bk')\times\\ 
  && \hspace{-1.5cm} |G_0(\bk,\omega)|^2 \int\Dx \, 
  \E^{-\I\bk\bx} \times \nonumber\\
  &&\hspace{-1.5cm}\left.\nabla^2_\br\right|_{\br=0}
  \underbrace{\big\langle V[h(\bx+\br)+l(\bx)-d] V[h(0)+l(0)-d]
    \big\rangle}_{\displaystyle\equiv C_2(\bx+\br,\bx;d)}. \nonumber 
\end{eqnarray}
Since the surface height profiles are Gaussian distributed, the pair
correlator $C_2$ will be a function of the first (which we have
chosen to be zero) and second moments of the distribution only. More
specifically, only the sum of the second moments enters the definition
of $C_2$, so we can define
\begin{equation}
  \label{c2_def}
  \bar C_2(K_h(\bx_1)+K_l(\bx_2);d) \equiv  C_2(\bx_1,\bx_2;d),
\end{equation}
with $K_h(\bx)=\langle h(\bx)h(0) \rangle$ and $K_l(\bx)=\langle
l(\bx)l(0) \rangle$, where the average is taken over the respective
height distributions. The exact form of $\bar C_2(K,d)$ is analyzed in
appendix~\ref{app_C2}. The main result of this analysis is that the
value of $\bar C_2$ is proportional to the normal load $N$ if
$|\bx_1|,|\bx_2|\lessim\sigma$. 
For the real space displacement correlation function follows (with
$\sigma \ll |\bx| \ll L$)
\begin{eqnarray}
  \label{rr_corr}
 &&\frac12 \langle\, [\br(\bx)-\br(0)]^2 \,\rangle =
 \int_\bk (1-\E^{\I\bk\bx}) 
 \langle \tilde \br_{\bk,0}\tilde \br_{-\bk,0}\rangle  \\
 &&=
 \int_{\bk,\bp} \!\!
 \frac{1-\E^{\I\bk\bx}}{\gamma^2 |\bk|^{2\alpha}} 
 \!\int \!\!\!\Dx_1\Dx_2 
 {\bf p}^2\E^{-\I\bk\bx_2+\I{\bf p}(\bx_2-\bx_1)}
 C_2(\bx_1,\bx_2;d) \nonumber \\
 && \sim    
  \sigma^2 \,{\cal NE}^2\, (\sigma/L)^D \times \rule{0mm}{5ex}
  \left\{ 
    \begin{array}{ll}
      \frac{1}{D_c-D}\,\left(\frac{|\bx|}{\sigma}\right)^{D_c-D} &
      \mbox{ for }D<D_c\\ 
      \ln(|\bx|/\sigma) & \mbox{ for }D=D_c, \rule{0mm}{3.5ex}
    \end{array}
  \right. \nonumber
\end{eqnarray}
where $\int_\bk$ is short hand for $\int\!\frac{d^Dk}{(2\pi)^D}$.
In the last equation, we have made use of
Eq.~(\ref{c2_N}) from App.~\ref{app_C2}, with $\kappa\approx 1$.
The (upper) critical dimension is given by
\begin{equation}
D_c=2\alpha\;,
\end{equation}
implying that the physically relevant situation $D=2$ and $\alpha=1$
corresponds to the marginal case.
Ignoring for a moment the dependence on length scales, the result is
proportional to ${\cal NE}^2 \sim NV_0/\gamma^2$; the linear
dependence on $N$ stems from replacing $V_0\E^{-d^2/4\hb^2}$ (being the
leading dependence on the mean distance $d$) by $N/L^D$ via
Eq.~(\ref{NormalLoad}). 

Like in other cases of collective pinning, one can now determine a
Larkin length $L_L$ from the condition
$\langle\,[\br(L_L)-\br(0)]^2\,\rangle \approx \sigma^2$
\cite{Larkin79}, i.e.\ $L_L$ is the length scale where typical
displacements become of the order of the corrugation length of the
substrate. This yields 
\begin{eqnarray}
  L_L &\approx& \sigma (L^D\epsilon/\sigma^D{\cal NE}^2)^{1/\epsilon}
  \nonumber \\
  &=&  \sigma[ \gamma^2\sigma^{2(2-\alpha)}/
  V_0\hb^{n+1}p\,]^{1/\epsilon},
\end{eqnarray}
where $\epsilon = D_c - D$.
Note that $L_L$ depends, apart from a combination of intrinsic model
parameters, only on the nominal pressure $p=N/L^D$.

What are the consequences for the static friction force, which in our
model is identified with the critical force at the depinning
transition \cite{NatterLeschhorn}?  $L_L$ will be smaller than the
system size only if we choose our model parameters such that ${\cal
N}\gg {\cal E}^{-2}(L/\sigma)^{D-\epsilon}$. In this case, it is
possible to estimate the static friction force in the standard way
\cite{LarkinKhmel,Feigelman,Bruinsma}: On small length scales, the
elastic energy is dominant compared to the interaction with the random
potential, so that adjacent sites move coherently. On larger length
scales, the random forces become more relevant until on the scale of
the Larkin length $L_L$, elastic and pinning forces are of the same
order of magnitude, so the elastic manifold becomes able to explore
the inhomogeneous force field emerging from the randomly distributed
contact points on scales larger than $L_L$. Regions of linear size
$L_L$ can hence be assumed to adapt independently to the disorder,
each of them giving an independent contribution to the pinning force
of the order of the elastic force on this scale. The force {\em
density} in a Larkin region is thus given by $\ffr({L_L}) \approx
\gamma \sigma L_L^{-\alpha}$, which is the product of a typical
gradient $\sigma/L_L^{2}$, a typical value of the elastic kernel
$\gamma L_L^{-D+2-\alpha}$, and the area $L_L^{D}$ of the region. The
total friction force is thus of the order 
\begin{equation}
\Ffr\approx(L/L_L)^D\,L_L^D\ffr({L_L})\approx \gamma L^D \sigma
L_L^{-\alpha},
\end{equation}
which is proportional to the nominal area $L^D$ and hence violates
Amontons's first law.  

With ${\cal N}(\sigma/L)^D\ll 1$ and ${\cal
E}\ll 1$, however, $L_L$ will typically be much larger than $L$. The
fact that in solid dry friction, the Larkin length is typically $\gg
L$, has been recently remarked also by Persson and Tossati
\cite{PerssonTossati} and by Caroli and Nozi{\`e}res
\cite{CaroliNozieres}. In the following, we will restrict ourselves
to this situation.  In the physical situation $D=2$ with bulk
elasticity $\alpha=1$, being just at the marginal dimension
$D_c$ (i.e.\ $\epsilon=0$), $L_L$ is {\em exponentially} large
\cite{LarkinKhmel,PerssonTossati}:
\begin{equation}
  L_L \sim \sigma \exp\left( c\,\frac{L^D}{\sigma^D{\cal NE}^2} \right),
\end{equation}
where $c$ is a dimensionless constant.

The elastic response of the surface also allows for a reduction of the
potential energy of the slider by decreasing the mean distance
$d$. Technically, this appears in the form of a correction factor to
the r.h.s.\ of Eq.~(\ref{NormalLoad}). It is calculated by expanding
$V'[h(\bx+\br(\bx))+l(\bx)-d]$ in $\br(\bx)$ and substituting for
$\br(\bx)$ the first order expression (\ref{r_firstorder}), yielding 
\begin{samepage}
\begin{eqnarray}
  && \langle V'[h(\bx+\br(\bx))+l(\bx)-d]\rangle = 
\langle  V'[h(0)+l(0)-d] \rangle  \nonumber  \\
  && {}+\int_\bk G_0(\bk,0) \int \Dx'\, \E^{-\I\bk\bx'} \times \nonumber \\
  && \quad\left.\nabla^2_\br\right|_{\br=0} 
  \langle V[h(\bx'+\br)+l(\bx')-d] V'[h(0)+l(0)-d] \rangle \nonumber \\
  && {}+ {\cal O}(V_0\hb^{n-1}{\cal E}^2). \rule{0cm}{2.5ex}
\end{eqnarray}
\end{samepage}
In this order of perturbation theory, the correction can be
represented by a factor $(1-c\,{\cal E}(\hb/d)^n)$ with another 
dimensionless constant $c$. Since this correction factor is smaller
than 1, the self-consistent determination of the mean distance $d$ via
Eq.~(\ref{NormalLoad}) will in turn lead to a slightly decreased value
of $d\rightarrow d-\Delta d$, so that the enlarged value of the
leading factor 
\begin{equation}
  \label{leading_factor_Deltad}
  \sim\E^{-(d-\Delta d)^2/4\hb^2}
\end{equation}
just compensates the
reduction. A similar correction in Eq.~(\ref{Coverage}), together with
the modified $d$, leads to a modified contact area $A_r\rightarrow
A_r+\delta A_r$.  

In order to determine the sign of the relative correction $\delta
A_r/A_r$, one has to compare the magnitude of the
relative corrections to $\langle V'(h+l-d)\rangle$ and to $A_r$, as
defined in Eq.~(\ref{def_Ar}), 
respectively. Using the result of Eq.~(\ref{c2}) in App.~\ref{app_C2},
one finds that the relative correction to $A_r$ is smaller by an
approximate factor of $\Gamma(n)\Gamma(n+1)/\Gamma(2n)$ (which is
smaller than 1 for $n>1$). Consequently,
the increase in $A_r$ due to $-\Delta d<0$ in the factor
Eq.~(\ref{leading_factor_Deltad}) will dominate, so the ratio $\delta
A_r/A_r$ (which is of order ${\cal E}$) is positive.

If there is a friction force at all,
it should depend on $\delta A$ and vanish for $\delta A=0$, because a
completely rigid surface is never pinned. We can hence give a
dimensional argument for a characteristic friction force
$\tilde\Ffr$. The simplest way to estimate $\tilde\Ffr$ is to write it
as the product of the excess contact area $\delta A_r\approx
\sigma^D{\cal{NE}}$ and a typical lateral force density
$V_0\hb^n/\sigma=\gamma\sigma^{1-\alpha}{\cal E}$
\begin{equation}
  \label{FfrEstimate}
  \tilde\Ffr \;\approx\; \gamma \sigma^{D+1-\alpha}{\cal NE}^2 .
\end{equation}
Here, a logarithmic correction of the order $\ln( L^D/{\cal N}
\sigma^D)$ has been omitted. Defining as usual the friction  
coefficient $\mu\equiv \Ffr/N$, this expression corresponds to a value
of $\mu$ of order ${\cal E}\hb/\sigma$ which depends on the ratio
between elastic and repulsive forces on small length scales but not on
the load. Below, we will show that this estimate gives indeed the
right order of magnitude of $\Ffr$.

\section{Kinetic and static friction}

Next, we consider the case of a moving slider, $v\equiv|{\bf v}|\neq
0$, driven by a 
finite force ${\bf F}$. The perturbative expansion is now set up in
\begin{equation}
{\bf u}(\bx,t) \equiv \br(\bx,t) - {\bf v}t,
\end{equation}
relative to
steady, rigid sliding.  For further simplification, we allow only for
displacements in the direction ${\bf e}_{\parallel}\equiv{\bf
  F}/|{\bf F}|$, corresponding to a further restriction
$\gamma_{\alpha\beta\gamma\delta}(\bx) = \gamma(\bx)\delta_{\alpha\gamma}
\delta_{\beta\delta}\delta_{\alpha 1}$ in Eq.~(\ref{GenElast}), where
$1$ denotes the direction  ${\bf e}_{\parallel}$. Hence, ${\bf
  u}(\bx,t)=u(\bx,t)\,{\bf e}_{\parallel}$. It has been argued in a 
closely related context (the depinning of a driven flux line in a
random medium) that this restriction will not alter the critical
dynamics in the average direction of motion \cite{Kardar}.  This
approximation will however overestimate the force needed to overcome a
repulsive trap, because asperities cannot avoid each other by simply
bending away. Recently it has been shown explicitly that this avoiding
process completely rules out {\em local} multistability in the case of
isolated isotropic traps, while multistability is re-established for 
anisotropic traps \cite{Tanguy}. 
Full account of $D$-dimensional lateral elasticity
has been taken in a simpler version of our model, where the random
potential energy is proportional to the product of two charge
densities on the two elastic manifolds, in \cite{Samokhin}; there,
however, the case of $L_L\ll L$ was considered.

To find the average dynamic friction force to lowest order in
perturbation theory, we follow a procedure used by Feigel'man for
driven interfaces \cite{Feigelman}. He calculated the first correction
to the mobility constant $\eta$, treating the average velocity $v$ as
a parameter which has to be determined self-consistently. Technically,
we proceed as follows: First, replace $\br(\bx,t)$ by ${\bf v}t + {\bf
  u}(\bx,t)$ in Eq.~(\ref{r_eom_fourier}), 
and expand the random potential part in this equation 
in powers of ${\bf u}(\bx,t)$.
Now, insert the Fourier transform of Eq.~(\ref{r_firstorder}), having
replaced the l.h.s.\ of this equation by $\tilde{\bf u}_{\bk,\omega}$
and the argument of $h$ on the r.h.s.\ by $\bx+{\bf v}t$,
into the next order expansion term in Eq.~(\ref{r_eom_fourier}),
\begin{equation}
  \int\!\!\Dx\,dt \,\E^{-\I\bk\bx-\I\omega t}
  u(\bx,t)
  \frac{\partial^2}{v^2\partial t^2} 
  V[h(\bx+{\bf v}t) + l(\bx) - d],
\end{equation}
and perform the average over $h$ and $l$. One finds to second
order in $V_0$
\begin{eqnarray}
\label{f_disorderaveraged}
  f &=& \frac{V_0^2\hb^{4n+1}}{d^{2n+1}}
       \frac{2^n\Gamma(2n+1)}{\sqrt{\pi}}
  \int_{\bp,\Omega} \!\!\!
  G_0({\bf p},\Omega) \times \\
  && \rule{0pt}{5ex}
  \int\!\!\Dx \,dt\, 
  \E^{\I{\bf px}+\I\Omega t} 
  \frac{1}{v^3}\frac{\partial^3}{\partial t^3}\,
  \Bigg[
    f_2\left(\frac{K_h(\bx+{\bf v}t)+K_l(\bx)}{2\hb^2}\right)\nonumber\\
  &&\hspace{2cm}\exp\left(-\,\frac{d^2}{2\hb^2+K_h(\bx+{\bf
       v}t)+K_l(\bx)}\right)\Bigg]. \nonumber 
\end{eqnarray}
Here, the representation of the random potential correlation function
obtained in App.\ \ref{app_C2}, with the dimensionless function
$f_2(\kappa)$ as defined in Eq.~(\ref{c2}), has been used.

The r.h.s.\ of Eq.~(\ref{f_disorderaveraged}),
multiplied with the full area $L_D$ of the slider, is identified 
with the total friction force $\Ffr(v)=L^D\ffr(v)$. The resulting
expression for $\Ffr(v)$ can be written in the form 
\begin{eqnarray}
\label{Full_Ffr}
&&\Ffr(v) = \\
 && L^D{\cal E}^2\gamma\sigma^{1-\alpha} \frac{v}{v_0}
  \int_{\fq,\fp}
  \frac{\fq_1^4}{|\fp|^{2\alpha} + (\frac{v}{v_0})^2\fq_1^2}
  \frac{\tilde c_2(\sigma\fq,\sigma(\fp-\fq))}{\sigma^{2D}}, \nonumber
\end{eqnarray}
where $\tilde c_2(\bp,\bp')$, defined in Eq.~(\ref{small_c2_tilde}) in
App.~\ref{app_C2}, is the Fourier transform of the random 
potential correlator defined in Eq.~(\ref{paircorr}). $\fp=\sigma\bp$
and $\fq=\sigma\bq$ are dimensionless integration variables, 
and we introduced the velocity scale
\begin{equation}
  \label{v0def}
  v_0\equiv\gamma\sigma^{1-\alpha}/\eta,
\end{equation}
which is a typical relaxation velocity.

Written in this way, the integral
on the r.h.s.\ of Eq.~(\ref{Full_Ffr}) is just a number, the
velocity dependence entering only via the ratio
$v/v_0$. Hence, we can  write the friction force in the form
\begin{equation}
  \label{Ffr}
  \Ffr(v) \approx \tilde\Ffr \;\phi\left(\frac{v}{v_0}\right),
\end{equation}
where $\phi(x)$ is a dimensionless
function that depends on the explicit form of
$V(z)$ and the statistics of the surface 
profiles. A detailed discussion of the velocity dependence in
$\phi(x)$ is given in App.~\ref{app_ffr}. Here, we will only
summarize the results.

At $v\approx v_0$, to begin with, $\Ffr(v)$ is
indeed found to be of the order $\tilde\Ffr$, which followed from our
na$\ddot{\i}$ve estimate Eq.\ (\ref{FfrEstimate}).  The amplitude
$\tilde\Ffr$, implying in particular the proportionality between
$\Ffr(v)$ and $N$, follows for similar reasons as those that led to
Eq.~(\ref{rr_corr}), making use of the random potential correlator
calculated in App.~\ref{app_C2}. Thus, our dimensional argument
(\ref{FfrEstimate}) is justified {\em a posteriori}.

In the high velocity regime $v\gg v_0$, the friction force decays
proportional to $1/v$, independent of $D$ and $\alpha$. Note that
as we have not included inertial terms in the equations of motion
(\ref{u_eom}) and (\ref{d_eom}), the experimentally observed velocity
strengthening behaviour of the friction force for high velocities can
not be reproduced by our model.  

In the most interesting regime $v\ll v_0$, the behaviour crucially
depends on the characterics of 
the surface profile correlator $k(x)$ and on the dimensionality.  If
the height-height correlator $k(x)$ is analytic in the origin, the
friction force reaches its maximum (of order 
$\tilde\Ffr$) at $v\approx v_0$, and it decreases for $v\ll v_0$ as
$(v/v_0)^{1-\epsilon/\alpha}$. This is the usual contribution to
friction from phononic damping, vanishing for $v\rightarrow 0$ if
$\epsilon/\alpha<1$. In the marginal dimension $D=2$, with bulk
elasticity, $\Ffr(v)$ thus depends linearly  on $v$ (times a logarithm
of $v/v_0$, cf.\ Eq.~(\ref{vel_dep}) in App.~\ref{app_ffr})
in the small velocity regime.

In order to find a finite static friction force, however, $\Ffr(v)$
has to bend towards a non-zero value for $v\rightarrow 0$.  In related
problems like the depinning of driven interfaces and charge density
waves, it has been shown that a finite pinning threshold appears due
to contributions to $\phi(x)$ on length scales larger than the Larkin
length \cite{NatterLeschhorn}. On these length scales, configurational
multistability emerges as a collective effect, leading to collective
pinning. This is reflected, in a renormalization group treatment, by
the renormalized random force correlator developing a cusp-like
singularity at the origin. Since in our case $L_L$ is typically larger
than the system size, the collective pinning mechanism is absent. We
thus have to find criteria for the existence of local multistability.

As mentioned above, real world surfaces often have self-affine
properties spanning several orders of magnitude. A (one dimensional)
surface profile, for instance, that is short range correlated on the
scale $\sigma$, and self-affine with a roughness exponent $\zeta=1/2$
below this scale, is described, in real space, by the correlator  
\begin{equation}
  K_h(x-x')=\langle h(x)h(x')\rangle =\hb^2 \E^{-|x-x'|/\sigma}.
\end{equation}
If the self-affinity covers the length scales from $\sigma$ down to a
microscopic length scale $a$, $\langle h(x)h(x')\rangle$ will
exhibit the cusp only when considering it on a coarser scale than $a$;
on finer scales, it is analytic again. In Fourier space, the
height-height correlator corresponding to such a profile is given by
\begin{equation}
  \label{KorrCuspFour}
  K_h(q) = \frac{2\,\E^{-(qa)^2}}{1+(\sigma q)^2}.
\end{equation}
A surface with this
correlation function can be generated by an Ornstein-Uhlenbeck process
\cite{Gardiner}: Given a stochastic spatial noise $\zeta(x)$ with
$\langle \zeta(x) \rangle=0$ and $\langle \zeta(x)\zeta(x') \rangle=
(2\hb^2/\sigma)\delta_a(x-x')$, where $\delta_a$ produces short range
correlations over a length scale $a$, the profile $h(x)$ obeying the
differential equation 
\begin{equation}
\frac{dh}{dx} = -\frac{1}{\sigma}\,h(x) + \zeta(x)
\end{equation}
has the desired statistical properties.

For such surface profiles, we find that $\phi(v/v_0)$ takes a finite
value of order 1 for $v_0(a/\sigma)^{1/\alpha}\ll v\ll v_0$. In this
case, the total friction 
force is indeed of the estimated order $\tilde\Ffr$ and almost
constant for $v<v_0$, and decays $\sim 1/v$ for $v\gg v_0$. More
generally, a finite static friction force is found if the
(unrenormalized) surface profile correlator $k(\bx)$ has a cusp in
the origin (precisely, we need $\lim_{x_\parallel\rightarrow
0_+}\partial_{x_\parallel}k(\bx) \neq 0$, where $x_\parallel$ is
the component of $\bx$ parallel to ${\bf f}$), which corresponds
to the first spatial derivative of $k(x)$ undergoing a jump (of order
1) at $x=0$. A surface characterized by such a correlator has local
slopes that may take arbitrarily high values, eventually leading to
multistability even for arbitrarily small roughness $\hb$.

Note that for $a\ll\sigma$, the lower velocity scale
$v_0(a/\sigma)^{1/\alpha}$
will typically be so small that it has no significance for a finite
sample, since velocity fluctuations will become so large that the
slider gets pinned. Consequently, for $L$ finite and hence in any
experimental situation the regime $v\ll v_0(a/\sigma)^{1/\alpha}$
where $\phi(v/v_0)\ll 1$ is unlikely to be observable.

\section{Strong pinning}

Our results have been obtained using the framework of perturbation
theory about weak disorder, which is sufficient in the weak pinning
limit and which is the natural starting point for a renormalization
group analysis that reveals configurational multistability on length
scales larger than $L_L$. In principle however, it is possible to have
multistability already locally on the scale of a single trap. This
situation corresponds to the strong pinning limit \cite{FukuyamaLee},
which cannot be treated successfully within finite order perturbation
theory.

Such a situation was recently considered by Caroli and Nozi{\`e}res
(CN) \cite{CaroliNozieres}.  They consider two flat surfaces with a
sparse distribution of bumps and sinks, where 'active' traps are
formed when two adjacent asperities are in contact. Writing the
interaction at an active trap as a potential energy $V(\rho)$ that
depends on the distance $\rho$ between their centers, they derive a
criterion for the existence of local (single-site) multistability
which reads 
\begin{equation}
\max|\partial^2_\rho V(\rho)|>E\sigma.
\end{equation}
$E$ is Young's modulus -- which, in $D=2$ and with $\alpha=1$, is our
$\gamma$ -- and 
$\sigma$ a typical length scale of the trap. Similar considerations
have been applied for instance in mean-field like descriptions of
driven interfaces or charge density waves \cite{Leschhorn}. The
frictional force is then proportional to the typical energy gap at a
spinodal jump, multiplied with the density of active traps.

To make contact with the CN model, we can identify the overlapping
asperities in our model (which are relatively few in the limit of
small normal load) with these active traps. We aim to derive an
estimate for the onset of local multistability in the case of short
range correlated surface profiles, with a typical corrugation length
$\sigma$, and a potential $V(z)=V_0 z^2\theta(z)$, in the physical
situation $D=2$ with bulk elasticity ($\alpha=1$). A typical energy
that can be stored in one active trap can roughly be estimated as
$V_0\hb^2\sigma^2$, and the second derivative with respect to a
lateral displacement will be of order $V_0\hb^2$. Comparing this with
the elastic energy term $\gamma\sigma$, we find as condition for the
presence of local multistability that 
\begin{equation}
{\cal E\ge O}(1),
\end{equation}
a condition which violates our assumptions.

This restriction can be relaxed to some extent if one looks for the
occurance of the {\em first} appearance of a multistable site in a
finite sample when 
tuning, for example, the potential amplitude $V_0$. The largest
asperity in a sample of linear size $L$ will have a height $h_{\max}$
of order $2\hb\ln^{1/2}((L/\sigma)^2)$, and with the expression for
the mean distance $d$ given after Eq.~(\ref{NormalLoad}) follows
$h_{\max}-d\approx \frac{2\hb^2}{d}\ln {\cal N}$. Using further that
the second derivative of a Gaussian correlated surface profile will
typically take a value of order $h_{\max}/\sigma^2$ at this maximum,
we arrive at the criterion 
\begin{equation}
{\cal E\ln N\ge O}(1)
\end{equation}
where $1\ll {\cal N}\ll (L/\sigma)^2$ has been used.

\section{Conclusion}

Expressions (\ref{FfrEstimate}) and (\ref{Ffr}) are the main results
of this paper. They describe a friction force which depends linearly
on the weight $N$ (up to logarithmic corrections), but not on the
nominal contact area $L^D$ and hence fulfills Amontons's laws. The
dependence on the potential parameter $n$ is weak. It is noteworthy
that this result was obtained in the limit $A_r\ll L^D$, where
statistics are dominated by {\em rare} events.

Trying other forms for the overlap interaction potential $V(z)$, it
turned out that the crucial ingredient leading to $\Ffr\sim N$ is the
cutoff below $z=0$. If one abandons this cutoff and uses, e.g., an
exponential potential $V(z)=V_0 \,\E^{z/z_0}$, the proportionality is
no longer valid; for the exponential potential, for example, one finds
instead to leading order that $\Ffr\sim N^2$.

To summarize, we have introduced a new stochastic model that
incorporates the interplay between bulk elasticity and surface
roughness in solid dry friction. To our knowledge, it is the first
purely elastic model that treats solid dry friction as a collective
phenomenon {\em and} reproduces the correct load dependence of the
friction force $\Ffr$, known as the Coulomb-Amontons's laws. For high
velocities $v$, $\Ffr$ decays like $1/v$, while the behaviour for
small $v$ depends on the surface profile statistics: For a smooth
surface, the static friction force vanishes in the weak pinning case,
while it is finite if the surface is characterized by a non-analytical
height-height correlator. We have also given an estimate for the
appearence of local multistability in the case of smooth interfaces.

We have not considered thermal effects in our study, because in most
situations where temperature changes do not strongly affect the
mechanical strengths of the sliding bodies, the friction coefficient
is found to be basically insensitive to temperature variations
\cite{Rabinovicz}. An example where friction does however strongly
depend on the temperature is the case of rubber sliding over hard
surfaces, where the friction properties are intimately connected with
the temperature dependent visco-elastic properties of the rubber
material \cite{Grosch}. In that specific case, the friction
coefficient $\eta$ in our model would become temperature dependent,
modifying the velocity scale $v_0$ (cf.\ Eq.~(\ref{v0def})) while
leaving the friction force amplitude $\tilde\Ffr$
(Eq.~(\ref{FfrEstimate})) constant. 

The present study can be extended in many directions. First, let us
shortly focus on the damping term in Eq.\ (\ref{u_eom}). In general,
also $\eta$ is non-local in space (and time): For long wavelengths, it
behaves like $\eta(k)\sim k^s$, where the precise value of $s$ depends
on the damping mechanism under consideration. Attenuation rates of
surface waves for a semi-infinite elastic body with a rough surface
lead to $s=3$ or 4 in the long wavelength regime \cite{Huang}. In this
case, the dynamics are dominated by a modified dynamical critical
dimension $D_{c,d}=2\alpha-s$, while the static properties are
unaffected.

In a situation where the Larkin length is smaller than the system
size, the perturbative results hold only in the limit of large
velocities; for this situation, the critical dynamics close to the
threshold force remain to be analyzed. Finally, inertial terms can be
included to properly describe the high velocity regime.

It is a pleasure to thank Jan Kierfeld, Tim Newman and Kirill Samokhin
for useful discussions. This work has been supported by the German
Israeli Foundation (GIF).

\begin{appendix}

\section{Random potential correlation function}
\label{app_C2}
Let $v_n(z) \equiv  \Theta(z) z^n$, cf.\ Eq.~(\ref{Vdef}). 
We want to calculate the pair correlator 
\begin{eqnarray}
\label{C2_def_app}
&&\bar C_2(K_h(\bx)+K_l(\bx');d) \equiv \\
&&\qquad\big\langle V[h(\bx)+l(\bx')-d] V[h(0)+l(0)-d] \big\rangle,
\nonumber 
\end{eqnarray}
cf.\ Eq.~(\ref{paircorr}), for fixed $\bx$
and $\bx'$. The first step is to rewrite Eq.~(\ref{C2_def_app}) using
the Fourier transform of $V(z)=\int_q \E^{\I qz} \tilde V_q$, followed
by performing the disorder average over the Gaussian height fields $h$
and $l$. This results in
\begin{equation}
  \int_{q,q'}\!\!\!\! \tilde V_q \tilde V_{q'} \E^{-\frac12
    (q^2+q'^2)(\hb^2+\lb^2)-qq'(K_h(\bx)+K_l(\bx')) - \I(q+q')d}.
\end{equation}
$V_q$ and $V_{q'}$ are Fourier transformed back to $V(z)$ and
$V(z')$, and the Gaussian integral over $q$ and $q'$ is carried out.
Before writing it down, we go over to dimensionless
functions and variables. 
We choose 
\begin{equation}
  l_0\equiv \sqrt{\langle h^2\rangle +\langle l^2\rangle}
\end{equation}
as the length scale in the direction perpendicular to the surface
plane. Let 
\begin{equation}
  \delta\equiv\frac{d}{l_0}, \qquad
  \kappa \equiv \frac{K_h(\bx) + K_l(\bx')}{l_0^2},
\end{equation}
so $\kappa$
ranges between 0 and 1, with $\kappa=1$ for $\bx=\bx'=0$ and $\kappa\rightarrow
0$ in the opposite limit $|\bx|,|\bx'| \gg \sigma$. Finally, let us define the
dimensionless function 
\begin{equation}
  c_2(\kappa,\delta)\equiv 
  \frac{\bar C_2(\kappa l_0^2,\delta l_0)}{V_0^2\hb^{2n}}.
\end{equation}
The resulting expression is then
\begin{eqnarray}
  \label{c2_zz}
  c_2(\kappa,\delta) &=& \int
  \frac{d\zeta d\zeta '\,v_n(\zeta)v_m(\zeta')}{2\pi\sqrt{1-\kappa^2}}\times \\
 &&\exp\left(  
  -\frac{(\zeta -\zeta ')^2+2(1-\kappa)(\zeta
  +\delta)(\zeta'+\delta)}{2(1-\kappa^2)}  
  \right). \nonumber
\end{eqnarray}
This is a general expression for arbitrary interaction
potential; from now on we will make use of the specific form of
$v_n(\zeta)$. 
Let us first consider the
limiting cases $\kappa=0$ and $\kappa=1$. First, for $\kappa=0$ one gets
\begin{eqnarray}
\label{kappa0}
  c_2(0,\delta) &=& \prod_{k\in\{m,n\}} 
    \int_0^\infty \frac{d\zeta }{\sqrt{2\pi}}\, \zeta^k \;\E^{-(\zeta
    +\delta)^2/2} \\
  &\stackrel{(\delta\gg1)}{\approx}&
  \frac{\Gamma(n+1)\Gamma(m+1)}{2\pi\;\delta^{m+n+2}}\;e^{-\delta^2}.
\end{eqnarray}
In the opposite limit $\kappa\rightarrow 1$, 
\begin{eqnarray}
\label{kappa1}
  c_2(1,\delta) &=& \int_0^\infty
  \frac{d\zeta }{\sqrt{2\pi}}   \;\zeta ^{m+n}\;\E^{-(\zeta +\delta)^2/2} \\
  &\stackrel{(\delta\gg1)}{\approx}&
  \frac{\Gamma(m+n+1)}{\sqrt{2\pi}\,\delta^{m+n+1}} \;\E^{-\delta^2/2}.
\end{eqnarray}
Comparing the $\delta$-dependence in eqs.~(\ref{kappa0}) and
(\ref{kappa1}) with the dependence of the mean distance $d$ on the
normal load, Eq.~(\ref{NormalLoad}), -- remember that we consider $d$
  to be a function of $N$ -- one finds to leading order
\begin{equation}
  \label{c2_N}
  c_2(\kappa,\delta) \sim 
  \left\{
    \begin{array}{ll}
      \left(\frac{\sigma}{L}\right)^D {\cal N} & \mbox{ for }
      \kappa\approx 1 \\ 
      \left(\left(\frac{\sigma}{L}\right)^D {\cal N}\right)^2 & 
      \mbox{ for } \kappa\approx 0, \rule{0ex}{3ex}\\
    \end{array}
  \right.
\end{equation}
where $\delta$ is to be considered a function of ${\cal N}/L^D$.

To make progress with Eq.~(\ref{c2_zz}) for arbitrary $\kappa$, one can 
introduce polar coordinates $\zeta \equiv r\cos\phi$, $\zeta '\equiv
r\sin\phi$. The $r$-integration can be carried out exactly, leaving
the one dimensional angular integral 
\begin{eqnarray}
\label{afterRint}
  c_2(\kappa,\delta) &=& \Gamma(\nu+1)\,(1-\kappa^2)^{\nu/2}
  \;\E^{-\frac{\delta^2}{1+\kappa}} \times \\
  &&\hspace{-0.7cm}\int_0^{\pi/2} \frac{d\phi}{2\pi}\;
  \frac{\cos^n\!\phi\;\sin^m\!\phi}{(1-\kappa\sin2\phi)^{\frac{\nu+1}{2}}}
  \;\E^{y^2(\phi)/4}D_{-(\nu+1)}(y(\phi)), \nonumber
\end{eqnarray}
where 
\begin{equation}
y^2(\phi)=\frac{1-\kappa}{1+\kappa} \,
\frac{1+\sin2\phi}{1-\kappa\sin2\phi} \, \delta^2,
\end{equation}
and $D_{-(\nu+1)}(y)$ with $\nu = m+n+1$
is a parabolic cylinder function. Equation~(\ref{afterRint}) can easily be
calculated numerically. For our present purpose it suffices however
to write
\begin{equation}
  \label{c2}
  c_2(\kappa,\delta) = 
  \frac{\Gamma(m+n+1)}{\sqrt{2\pi}\,\delta^{m+n+1}}\;
  f_2(\kappa)\;\E^{-\delta^2/(1+\kappa)},  
\end{equation}
where  $f_2(\kappa)$ only weakly depends on $\kappa$ ('weakly' compared
to the leading exponential dependence). It is a monotonically
growing function with
\begin{equation}
  f_2(\kappa) \approx 
    \left\{
      \begin{array}{cl}
        \frac{\Gamma(n+1)\Gamma(m+1)}{\Gamma(m+n+1)}\frac{1}{\sqrt{2\pi}\delta}         & \mbox{ for } \kappa = 0 \\
        1 & \mbox{ for } \kappa = 1. \rule{0cm}{2.5ex}
      \end{array}
    \right.
\end{equation}
For large enough $\delta$, we can neglect the dependence on
$f_2(\kappa)$: The ratio $\E^{-\delta^2/2}/f_2(0)$ is $\approx 1/50$
for $\delta=4$, $\approx 3\times 10^{-4}$ for $\delta=5$ etc., with
$n=m=2$. 

Since we will primarily need the Fourier transform of the potential
correlation function, we define
\begin{equation}
  \label{C_pp_def}
  \tilde C_2(\bp,\bp')\equiv \int\Dx\,\Dx'\;
  \E^{\I\bp\bx+\I\bp'\bx'}\,C_2(\bx,\bx')
\end{equation}
and analyze its behaviour for large $\delta$ such that
$\E^{-\delta^2/2}\ll 1$. Under this constraint, it is a good
approximation to consider only the leading $\delta$- and
$\kappa$-dependence of $c_2(\kappa,\delta)$,
i.e. $c_2(\kappa,\delta)\simeq \E^{-\delta^2/(1+\kappa)}$, making 
an error of ${\cal O}(1)$. For sake of simplicity, let us assume
henceforth that both surface profiles obey the same distribution
$K_h(\bx)$ with
\begin{equation}
  \frac{K_h(\bx)}{l_0^2} \approx 
  \left\{
    \begin{array}{ll}
      1/2 & \mbox{ for } |\bx| \lessim \sigma \\
      0 & \mbox{ for } |\bx| \gg \sigma . \rule{0cm}{2.5ex}
    \end{array}
  \right.  
\end{equation}
Let 
\begin{equation}
\label{small_c2_tilde}
  \tilde c_2(\bp,\bp')\equiv \tilde C_2(\bp,\bp')\,/\,V_0^2\hb^{2n}.
\end{equation}
Because $C_2(\bx,\bx')$ decays to a nonzero constant for
$|\bx|\gg\sigma$, it is convenient to write $\tilde c_2(\bp,\bp')$ as
the sum of four terms:
\begin{equation}
  \label{four_terms}
  \tilde c_2^\alpha\delta_{\bp}\delta_{\bp'} + 
  \tilde c_2^\beta(\bp)\delta_{\bp'} +
  \tilde c_2^\beta(\bp')\delta_{\bp} +
  \tilde c_2^\gamma(\bp,\bp'),
\end{equation}
only the last two of which will contribute in the integrals $\tilde
C_2$ appears in (cf.\ for example Eq.~(\ref{app_fdef})). The other
two terms will be eliminated due to the $\delta_\bp$--term being
multiplied with a power of $p_1$.

Now, first consider Gaussian correlated surface profiles
$k(x)=\E^{-x^2/2}$ (cf. the definition in
Eq.~(\ref{def_kx})). Performing a saddle-point expansion about small
$\bx$ and $\bx'$ in $C_2(\bx,\bx')$ and then Fourier transforming,
$\tilde c_2^\beta(\bp)$ is found to be proportional to
$(\sigma/\delta)^D\E^{-2\delta^2/3}$ for $|\bp|\lessim \delta/\sigma$
and exponentially damped for larger $\bp$. Correspondingly,
\mbox{$\tilde{c}_2^\gamma(\bp,\bp')\sim(\sigma/\delta)^{2D}\E^{-\delta^2/2}$}
for 
$|\bp|,|\bp'|\lessim \delta/\sigma$, with exponential suppression for
$|\bp|$ or $|\bp'|\morsim \delta/\sigma$.

If the surface correlation function is taken to have a cusp in the
$x_1$-direction, the only -- and important -- difference arises in the
large $p_1$-behaviour of $\tilde c_2^\beta$ and $\tilde c_2^\gamma$. With the
modification that 
\begin{equation}
\label{k_x1_cusp}
  k(x_1)=\E^{-|x_1|},
\end{equation}
leaving the dependence on the
perpendicular spatial coordinates unmodified, we find
\begin{equation}
\tilde c_2^\beta(p_1)\sim \sigma_b/(1+p_1^2\sigma_b^2)
\end{equation}
with $\sigma_b=9\sigma/2\delta^2$, and a similar result for $\tilde
c_2^\gamma(p_1,p_1')$. In the perpendicular direction $\bp_\perp$,
$c_2^\beta$ and $c_2^\gamma$ are again exponentially suppressed.

\section{Dynamic friction force}
\label{app_ffr}
We want to analyze the expression (\ref{f_disorderaveraged}) that
gives the kinetic friction force density at velocity $v$ in first
non-vanishing order perturbation theory. Two generic scenarios will be
considered, assuming in both cases that the two surface profiles obey
identical distributions. First, we consider the case of short range
correlated profiles, $k(x)=\E^{-x^2/2}$, and second,
profiles that are self-affine in the direction of ${\bf F}$. Let us
rewrite Eq. (\ref{f_disorderaveraged}) as 
\begin{equation}
  \label{app_fdef}
  f = \int_{\bp} \int_{\bq}
  \;G_0(\bp,q_1 v)\;\I q_1^3\; \tilde C_2(\bq,\bp-\bq),
\end{equation}
with $\tilde C_2(\bp,\bp')$ as defined in Eq.~(\ref{C_pp_def}).
Only the imaginary part of the propagator
\begin{equation}
  \label{ImG0}
  \Im \left( G_0(\bp,q_1 v) \right) = 
  -\frac{\eta q_1 v }{ \gamma^2 |\bp|^{2\alpha}+(\eta q_1 v)^2 }
\end{equation}
gives a contribution to the integral, the integral over the real part
vanishes by symmetry. Now, $\tilde C_2(\bp,\bp')$ is of order 
\begin{equation}
  V_0^2\hb^{2n}\E^{-\delta^2/2} \sim 
  \frac{\gamma \sigma^{3-\alpha}}{L^D} \;\tilde\Ffr
  \quad\mbox{ for } 1/L < |\bp^({}'^)| < 1/\sigma,
\end{equation}
with $\delta$ as defined in App.~\ref{app_C2} 
and $\tilde\Ffr$ from Eq.~(\ref{FfrEstimate}),
and vanishes rapidly for $|\bp^({}'^)|>1/\sigma$. This is the contribution
from $\tilde c_2^\gamma(\bp,\bp')$, cf.\ Eq.~(\ref{four_terms}). The second
term in the denominator in (\ref{ImG0}) serves as an infrared cutoff
for the $\bp$-integration at $p\approx \sigma^{-1}(v/v_0)^{1/\alpha}$,
giving rise to a velocity dependence
\begin{equation}
 \label{vel_dep}
 \phi(v/v_0) \sim \left\{
   \begin{array}{cl}
     \frac{1}{D-2\alpha}\left(\frac{\DS v}{\DS v_0}\right)^{D/\alpha-1}
     &\mbox{ for } D<2\alpha\\ 
     \frac{\DS v}{\DS v_0}\ln(v/v_0)& \mbox{ for } D=2\alpha, \rule{0cm}{3ex}
   \end{array}
 \right.
\end{equation}
hence confirming the discussion after Eq.~(\ref{Ffr}) for
velocities $v\ll v_0$. For $v\gg v_0$, the $1/v$-behaviour follows from
the dominance of the second term in the denominator of
Eq.~(\ref{ImG0}), regardless of the dimension $D$ and the value of
$\alpha$. The term $\tilde c_2^\beta(\bq)\delta_{(\bp-\bq)}$ gives an
independent contribution to the integral in Eq.~(\ref{app_fdef})
proportional to 
\begin{equation}
  \label{IntC2Beta}
  \int_{\bp} \frac{\I p_1^3}{\gamma|\bp|^\alpha +\I \eta vp_1}
  \;\tilde c_2^\beta(\bp).
\end{equation}
For small $v\ll v_0$, it is linear in $v$, and, apart from the velocity
dependence, it is smaller by a factor of $\E^{-\delta^2/6}$ compared
to the leading term, so we can neglect it here.

Now consider surface profiles that are characterized by a
correlation function modified via Eq.~(\ref{k_x1_cusp}),
having a cusp in the direction of the applied force, and being
analytic in directions perpendicular to it.
In contrast to the previous situation, $\tilde C_2(\bq,\bp-\bq)$ is 
exponentially damped now only in the direction perpendicular to ${\bf
  F}$, while the large $p$-behaviour of $c_2^\gamma$ in the sliding
direction is $\sim(p_1-q_1)^{-2}q_1^{-2}$, and $\sim p_1^{-2}$ for
$c_2^\beta$. In the contribution to Eq.~(\ref{app_fdef}) containing
the term $c_2^\gamma$, performing the $q_1$-integration leads to 
\begin{equation}
  \frac{\pi}{2\gamma\sigma^4}
  \int_{\bp,\bq_\perp} \frac{2|\bp|^\alpha+\tilde
  v}{(|\bp|^\alpha+\tilde v)^2}\;
  \tilde C_2(\bq_\perp,\bp_\perp-\bq_\perp),
\end{equation}
where $\tilde v = v/v_0\sigma^\alpha$.
The remaining integrals can now easily be carried out, yielding a result
which, provided that $D>\alpha$, is {\em independent} of $v$ for $v\ll
v_0$. It is of order
\begin{equation}
  \frac{V_0^2\hb^{2n}\E^{-\delta^2/2}}{(D-\alpha)\gamma\sigma^{3-\alpha}} 
  \approx \frac{1}{(D-\alpha)\,L^D} \;\tilde\Ffr.
\end{equation}
For $\alpha=1$, this result is multiplied with a factor of $\ln(\sigma /
a)$, where $1/a$ is the UV-cutoff of the $p_1$-integra\-tion, $a$
setting the smallest length scale down to which the self-affinity of
the surface profile holds, cf.\ Eq.~(\ref{KorrCuspFour}). The result
for the friction force in this case is independent of $v$ in the regime
$(a/\sigma)^{1/\alpha}v_0\ll v \ll v_0$. For even lower velocities,
the fact that the correlator in Eq.~(\ref{KorrCuspFour}) is analytic
on length scales smaller than $a$ comes into play again, leading to
the same $v$-dependence of $\phi(v/v_0)$ in this regime as given by
Eq.~(\ref{vel_dep}). 

Again, there is an independent contribution from 
$\tilde c_2^\beta(\bq)\delta_{(\bp-\bq)}$ of the form Eq.~(\ref{IntC2Beta}).
This term has a dependence on
the mean distance $\sim\E^{-2\delta^2/3}$ and is hence 
proportional to $((\sigma/L)^D{\cal N})^{4/3}$. For $\alpha=2$, this
contribution is negligible, but for $\alpha=1$, it suffers from a 
linear divergence in the UV-cutoff $\sim \sigma/a$ due to the large
$p_1$-behaviour of $\tilde c_2^\beta$. We have checked
that in higher orders perturbation theory no divergences in higher
order of $\sigma/a$ appear, so this divergence will be compensated by
the $\delta$-dependence of this term for moderately large $\delta$.  
\end{appendix}

\end{multicols}
\end{document}